\documentclass[12pt, preprint]{aastex}

\newcommand\zp{$\zeta$~Pup}
\newcommand\zo{$\zeta$~Ori}
\newcommand\lya{Ly$\alpha$}
\newcommand\hea{He$\alpha$}
\newcommand\lyb{Ly$\beta$}
\newcommand\bsob{\ensuremath{\beta_{\mathrm{Sob}}}}
\newcommand\tss{\ensuremath{\tau_{0,*}}}
\newcommand\pavg{\ensuremath{\left < p \right >}}
\newcommand\xmm{{\it XMM-Newton}}

\begin{document}

\title{Evidence for the importance of resonance scattering in X-ray emission
  line profiles of the O star $\zeta$~Puppis}

\author{Maurice A.\,Leutenegger\altaffilmark{1}, Stanley
  P. Owocki\altaffilmark{2}, Steven M. Kahn\altaffilmark{3}, Frits
  B. S. Paerels\altaffilmark{1}} 
\altaffiltext{1}{Department of Physics and Columbia Astrophysics Laboratory,
  Columbia University, 550 West 120th Street, New York, NY 10027;
  \email{maurice@astro.columbia.edu}}  
\altaffiltext{2}{Bartol Research Institute, University of Delaware, 217 Sharp
  Laboratory, Newark, DE 19716} 
\altaffiltext{3}{Kavli Institute for Particle Astrophysics and Cosmology,
  Stanford Linear Accelerator Center and Stanford University, 2575 Sand Hill
  Road, Menlo Park, CA 94025} 

\shorttitle{Resonance scattering in \zp}
\shortauthors{Leutenegger et al.}

\begin{abstract}

We fit the Doppler profiles of the He-like triplet complexes of \ion{O}{7} and
\ion{N}{6} in the X-ray spectrum of the O star \zp, using \xmm\, RGS
data collected over $\sim 400$ ks of exposure. We find that they cannot be
well fit if the resonance and intercombination lines are constrained to have
the same profile shape. However, a significantly better fit is achieved with a
model incorporating the effects of resonance scattering, which causes the
resonance line to become more symmetric than the intercombination line for a
given characteristic continuum optical depth $\tau_*$. We discuss the
plausibility of this hypothesis, as well as its significance for our
understanding of Doppler profiles of X-ray emission lines in O stars.

\end{abstract}

\keywords{stars: early type --- star: winds, outflows --- techniques:
  spectroscopic --- stars: individual (\zp)}

\section{Introduction}

High resolution X-ray spectra obtained with diffraction grating spectrometers
on the {\it Chandra} and \xmm\, X-ray observatories have
revolutionized our understanding of the X-ray emission of O stars in the last
five years. In the canonical picture, the X-rays are emitted in plasmas heated
by shocks distributed throughout the wind \citep[][]{CS83, Cor93,
  Corcoranetal1994, Hel93}; the shocks are created by instabilities in the
radiative driving force \citep[e.g.][]{LW80, OCR88, Cooper94, FPP97}. Although
some stars show anomalous X-ray emission that can be explained by a hybrid
mechanism involving winds channelled by magnetic fields \citep[e.g. $\tau$~Sco
and $\theta^1$~Ori~C,][]{Del02, Del06, Cel03, Gel05}, a number of 
``normal'' O stars have X-ray spectra that are mostly consistent with the
wind-shock paradigm (e.g. \zp, \zo, and $\delta$~Ori). The papers describing
the first few high resolution spectra of normal O stars obtained reported some
inconsistencies with expectations \citep{WC01, Kel01, CMWMC01, MCWMC02,
  Wel04}, but more recent quantitative work based on the simple empirical
Doppler profile model of \citet[][hereafter OC01]{OC01} has resolved many of
these problems (\citealt{KCO03}; \citealt{Cel06, LPKC06}). The main
outstanding problem is the relative lack of asymmetry in emission line Doppler
profiles, which, if taken at face value, would imply reductions in the
literature mass-loss rates of an order of magnitude \citep{KCO03, Cel06, OC06}.

Although there is mounting evidence from other lines of inquiry suggesting
that the literature mass-loss rates may be systematically too high
(\citealt{MFSH03, Hel03}; \citealt{BLH05, FMP06}), there are also subtle
radiative transfer effects that could cause emission line profiles to be more
symmetric than one might naively expect. Two effects that have been
investigated in the literature are porosity \citep{FOH03, OFH04, OFH06, OC06}
and resonance scattering \citep[][hereafter IG02]{IG02}. Porosity could lower
the effective opacity of the wind to X-rays, thus symmetrizing emission
lines. However, \citet{OFH06} and \citet{OC06} have found that the
characteristic separation scale of clumps must be very large to show an
appreciable effect on line profile shapes, which makes it difficult to achieve
a significant porosity effect. Resonance scattering can symmetrize Doppler
profiles by favoring lateral over radial escape of photons; it is an
intriguing possibility, but to date it has not been tested experimentally.

In this paper, we present evidence for the importance of resonance
scattering in some of the X-ray emission lines in the spectrum of the O star
\zp. We show that the blend of resonance and intercombination lines of
two helium-like triplets in the very high signal-to-noise ratio \xmm\,
Reflection Grating Spectrometer (RGS) spectrum of \zp\ cannot be well fit by
assuming that both lines have the same profile but can be much better fit by
assuming the profile of the resonance line is symmetrized by resonance
scattering.
 
This paper is organized as follows: in \S~\ref{sec:data} we discuss the
reduction of over 400 ks of \xmm\, RGS exposure on \zp; in \S~\ref{sec:nr} 
we briefly recapitulate the results of OC01 and \citet{LPKC06} for Doppler
profile modelling (\S~\ref{sec:nrmodel}), and we show that the He-like OC01
profile model does not give a good fit to the \ion{O}{7} and \ion{N}{6}
triplets of \zp\ (\S~\ref{sec:nrfit}); in \S~\ref{sec:rs} we generalize the
results of OC01 to include the effects of resonance scattering as derived in
IG02 (\S~\ref{sec:rsmodel}), and we fit this model to the data
(\S~\ref{sec:rsfit}); in \S~\ref{sec:discussion} we discuss our results; and
in \S~\ref{sec:conclusions} we give our conclusions. 

\section{Data reduction \label{sec:data}}

The data were acquired in 11 separate pointings. The first two observations
were performance verification, while the rest were calibration; they are all
available in the public archive. The ODFs were processed with SAS version
7.0.0 using standard procedures; periods of high background were filtered
out. Only RGS \citep{dHel01} data were used in this paper, but EPIC data are
available for most of the observations. Processing resulted in a co-added total
of 415 ks of exposure in RGS1 and 412 ks in RGS2. The observation IDs used
and net exposure times are given in Table~\ref{tab:obslog}.

RGS has random systematic wavelength scale errors with a $1\sigma$ value of
$\pm 7$ m\AA\ \citep{dHel01}. A 7 m\AA\ shift could lead to significant
systematic errors in the model parameters measured from a line
profile. Because of this, we co-add all observations using the SAS task {\tt
  rgscombine}. Assuming the systematic shifts are randomly distributed,
co-adding the data will result in a spectrum that is almost unshifted
(depending on the particular distribution of shifts of the individual
observations), but that is broadened by 7 m\AA; this effect is much easier to
account for in our analysis. We have assumed that the data do not vary
intrinsically. We have not formally verified that the data show no significant
intrinsic variation, but upon visual inspection the data do not appear to vary
more than expected from statistical fluctuations combined with the
aforementioned random systematic errors in the wavelength scale.

Spectral fitting was done with XSPEC version 12.2.1; the line profile models
are implemented as local models. The $C$ statistic \citep{C79} is used instead
of $\chi^2$ because of the low number of counts per bin in the wings of the
profiles. 

Because of the failed CCD on RGS2, we only have RGS1 data for \ion{O}{7}
\hea. We only fit RGS2 data for \ion{N}{6} \hea\ because the 
complex falls on a chip gap for RGS1. 

For each complex we fit, we first measured a local continuum strength
from a nearby part of the spectrum uncontaminated by spectral features. We
modeled this continuum as a power law with an index of 2, which is flat when
plotted against wavelength. When fitting a line profile, we fit a combination
of the local continuum (fixed to the measured value) plus the line profile
model to the data. 

For the \ion{N}{6} \hea\ complex, we also included emission from
\ion{C}{6} \lyb\ at 28.4656 \AA, since the red wing of this line
overlaps the blue wing of the resonance line of \ion{N}{6} \hea. The
model parameters for \ion{C}{6} \lyb\ are assumed to be the same as
for \ion{N}{6}, and it is assumed to be optically thin to resonance
scattering.

Emission lines and complexes were fit over a wavelength range of 
$\lambda_{-} < \lambda < \lambda_{+}$. Here $\lambda_\pm = \lambda_0 (1
\pm v_\infty / c) \pm \Delta \lambda$, where $\Delta \lambda$ is the
resolution of RGS at that wavelength. $\lambda_0$ is the shortest wavelength
in the complex for $\lambda_{-}$ and the longest wavelength for
$\lambda_{+}$. 


\section{Best fit He-like profile model \label{sec:nr}}

\subsection{The profile model}
\label{sec:nrmodel}

In this section we briefly recapitulate the results of OC01 for the Doppler
profile of an X-ray emission line from an O star wind and the extension of
these results to a He-like triplet complex by \citet{LPKC06}.

In the physical picture of the OC01 model, the wind is a two-component fluid;
the bulk of the wind is relatively cool material of the order of the
photospheric temperature, while a small fraction of the wind is at
temperatures of order 1--5 MK, so that it emits X-rays. The cool part of the
wind has some continuum opacity to X-rays and can absorb them as they leave
the wind. 

The OC01 formalism casts the line profile in terms of a volume integral over the
emissivity, attenuated by continuum absorption:
\begin{equation}
L_\lambda = 4\pi \int dV \eta_\lambda(\mu, r) e^{-\tau_c(\mu, r)}
\end{equation}
where $\eta_\lambda(\mu, r)$ is the emissivity at the observed wavelength
$\lambda$ and $\tau_c(\mu, r)$ is the continuum optical depth to X-rays of
the wind. 

The line profile can be expressed in terms of the scaled wavelength
$x \equiv (\lambda / \lambda_0 - 1) c / v_\infty = - v_z / v_\infty$; this
gives the shift from line center in the observer's frame in units of the wind
terminal velocity. The sign convention is such that positive $x$ corresponds
to a redshift. 

OC01 derive an expression for the line profile in terms of an integral over the
inverse radial coordinate $u = R_* / r$ [cf. their eqn.~(9)]:
\begin{equation}
  L_x = L_0 \int_{0}^{u_x}\, du\, \left. \frac{f_X(u)}{w^3(u)} e^{-\tau(\mu,u)} 
  \right|_{\mu = -x/w(u)}.
\label{eqn:lx}
\end{equation}
In this equation we have used the following expressions: $w(u) \equiv v(u) /
v_{\infty} = (1 - u)^{\beta}$ is the scaled velocity, $\tau(x,u)$ is the
(continuum) optical depth to X-rays emitted along a ray to the observer, 
$f_X(u) \propto u^q$ is the filling factor of X-ray emitting plasma, and
$u_x \equiv \min(u_0, 1 - |x|^{1/\beta})$ is the upper limit to the
integral; $u_0 = R_* / R_0$ is the inverse of the minimum radius of X-ray
emission $R_0$, and $1 - |x|^{1/\beta}$ is a geometrical cutoff for the
minimum radius that emits for a given value of $x$. The integral for $L_x$ can
be evaluated numerically.  

The optical depth in this expression is derived in OC01. It is written as the
product of the characteristic optical depth $\tau_* = \kappa \dot{M} / 4 \pi
R_* v_\infty$ times a dimensionless integral containing only terms depending
on the geometry. It can be evaluated analytically for integer values of
$\beta$. For non-integer values of $\beta$, the optical depth must be
evaluated numerically, which is computationally costly, and thus not
convenient in conjunction with the radial integral of the line
profile. Because of this we assume $\beta = 1$ throughout this paper,
which is a good approximation for \zp, as well as for O stars in general.

The interesting free parameters of this model are the exponent of the radial
dependence of the X-ray filling factor, $q$; the characteristic optical depth
to X-rays of the cold plasma, $\tau_*$; and the minimum radius for the onset
of X-ray emission $R_0$.

\citet{LPKC06} extend this analysis to a He-like triplet complex. The only
difference is that the forbidden-to-intercombination line ratio has a radial
dependence due to photoexcitation of the metastable upper level of the
forbidden line:
\begin{equation}
{\cal R} \equiv \frac{f}{i} = {\cal R}_0 \frac{1}{1 + \phi / \phi_c} 
= {\cal R}_0 \frac{1}{1 + 2 P W(r)} .
\end{equation}
Here $\phi$ is the photoexcitation rate from the upper level of the forbidden
line; it depends on the photospheric UV flux and scales with the geometrical
dilution $W(r)$; $\phi_*$ is the photoexcitation rate near the photosphere,
so that $\phi = 2 \phi_* W(r)$; $\phi_c$ is the critical photoexcitation rate,
which is a parameter of the ion; and $P = \phi_* / \phi_c$ is a convenient
dimensionless parameter that gives the relative strength of photoexcitation
and decay to ground near the star such that ${\cal R}(R_*) = {\cal R}_0 / (1 +
P)$. In this paper, we use values of $P$ calculated from TLUSTY stellar
atmosphere models \citep{LH03} as described in \citet{LPKC06}. Values of
${\cal R}_0$ are taken from \citet{PMDRK01}. 

To modify the expressions for the forbidden and intercombination line profiles
to account for this effect, the emissivity is multiplied by the normalized
line ratio: 
\begin{equation}
\eta_f(r) = \eta(r) \frac{{\cal R}(r)}{1 + {\cal R}(r)}
\end{equation}
\begin{equation}
\eta_i(r) = \eta(r) \frac{1}{1 + {\cal R}(r)}.
\end{equation}

\subsection{Best fit model}
\label{sec:nrfit}
In this section we model the Doppler profiles of the \ion{O}{7} and
\ion{N}{6} He-like triplets with the He-like profile of \citet{LPKC06}
described in \S~\ref{sec:nrmodel}. The forbidden line is very weak for these
two ions, and the intercombination line profile predicted by the model is not
very different from the resonance line profile. The main difference in the
profile of the resonance and intercombination lines is that the extremes of
the wings are somewhat weaker. This is because the $f/i$ ratio reverts to the
low UV flux limit at very large radii ($> 100 R_*$ for \ion{O}{8} for \zp),
so the intercombination line strength is reduced by a factor of a
few. However, this has only a small effect on the profile shape.

Although it is weak, the predicted strength of the forbidden line is a good
check on the consistency of the profile model. The value of the characteristic
optical depth $\tau_*$ can have a strong effect on the observed $f/i$ ratio
by setting the value of $R_1$, the radius of optical depth unity. However,
this effect is degenerate with the value of $q$, the exponent of the radial
dependence of the X-ray filling factor.

\begin{figure}[p]
  \begin{center}
    \plotone{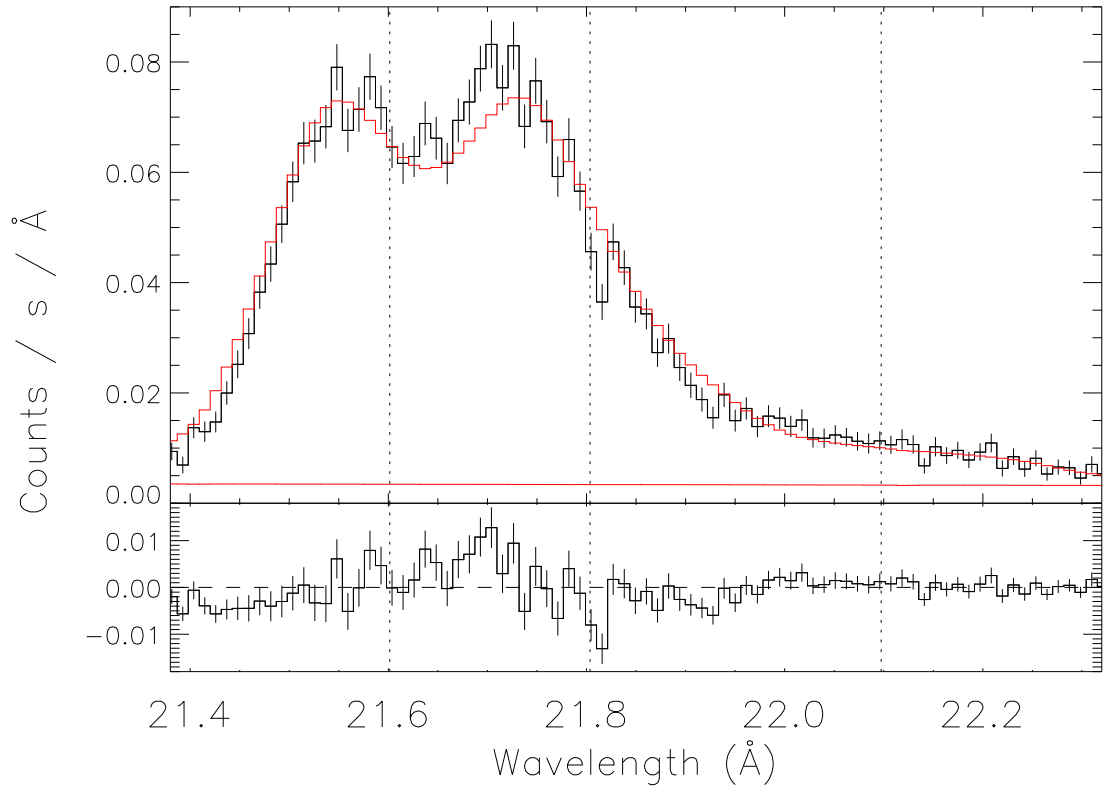}
  \end{center}
\caption{\ion{O}{7} triplet with best-fit OC01 He-like triplet model (not
  including the effects of resonance scattering). The top panel shows the data
  in black with error bars and the model in red. The flat red line shows the
  assumed continuum strength. The bottom panel shows the fit residuals.}
\label{fig:o7nors}
\end{figure}

\begin{figure}[p]
  \begin{center}
    \plotone{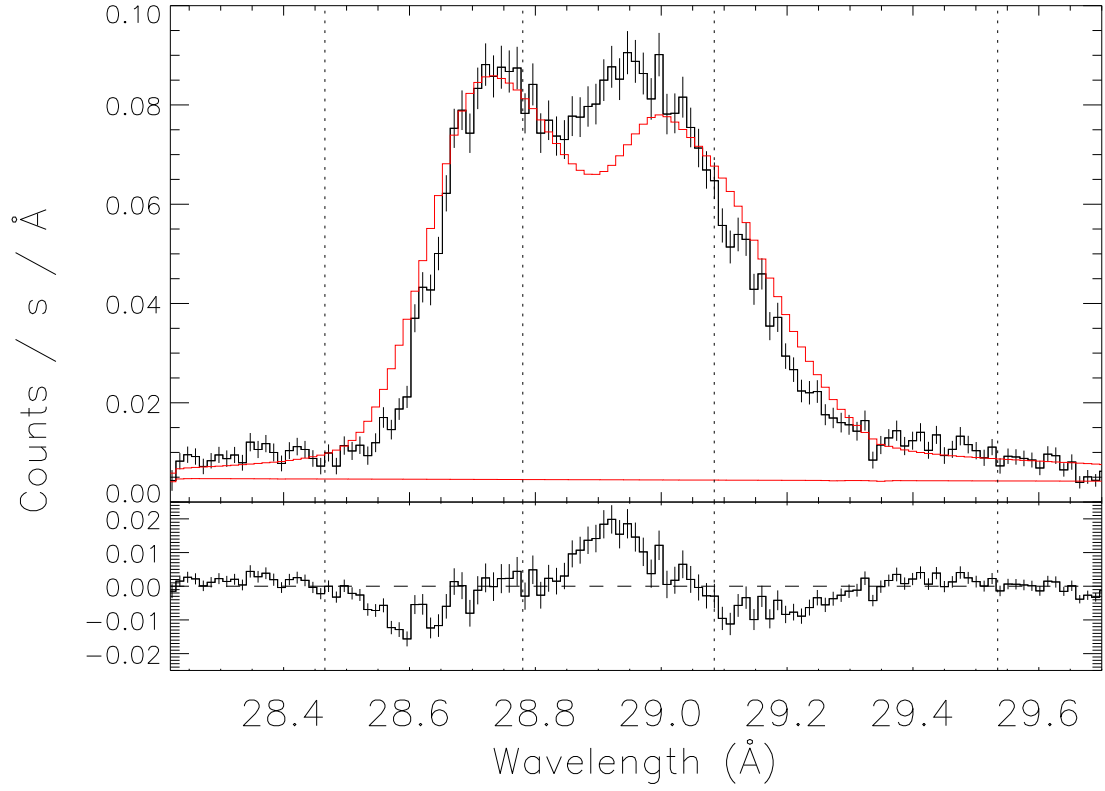}
  \end{center}
\caption{Same as Fig.~\ref{fig:o7nors}, but for the \ion{N}{6} triplet.
  The \ion{C}{6} \lya\ line at 28.4656 \AA\ is also included in the fit, as
  well as the other fits to the \ion{N}{6} triplet.}
\label{fig:n6nors}
\end{figure}

In Figures~\ref{fig:o7nors} and \ref{fig:n6nors}, we show the Doppler profiles
of the \ion{O}{7} and \ion{N}{6} He-like triplets, together with the best-fit
models. The best-fit parameters are given in Tables~\ref{tab:O7} and
\ref{tab:N6}. There are significant residuals in both fits. The \ion{N}{6}
triplet shows stronger residuals than \ion{O}{7}. The residuals have a
systematic shape: the model predicts a greater flux than the data on the blue
wing of the resonance line and the red wing of the intercombination line,
while it underpredicts the data in the center of the blend.

The systematic nature of the residuals shows that the shapes of the Doppler
profiles of the resonance and intercombination lines are different. The
residuals are consistent with the model resonance line being too blue and
therefore too asymmetric, and the model intercombination line being too red
and therefore too symmetric.

Resonance scattering has been proposed by IG02 as an explanation for
the properties of O star X-ray emission line Doppler profiles. If it is
important, it can cause significant symmetrization of profiles of strong
resonance lines. Because this is in qualitative agreement with our
observations, we explore this idea further. 

\section{Best-fit model including the effects of resonance
  scattering \label{sec:rs}} 

\subsection{Incorporating resonance scattering into OC01 \label{sec:rsmodel}}

The wind structure that gives rise to X-ray emission is likely to be 
extremely complex, with a highly nonmonotonic velocity field coupled to 
strong density variations in all three dimensions.
Nonetheless, even in such a medium, the overall flow pattern, to
first approximation, should still be largely dominated by the radial 
acceleration of a mostly radial, supersonic outflow.
In such a medium, the radiative transfer of line emission can be well 
modelled in terms of localized escape methods first developed by
\citet{Sobolev60}, so we use this as our basis for investigating
the possible role of resonance scattering in X-ray emission lines.
Our approach here generally follows that taken by IG02, with two
modest generalizations:
(1) Instead of assuming a fully optically thick
line, we allow the line optical depth to be a free parameter that can 
range from the optically thin to thick limits; and
(2) instead of assuming a constant-speed expansion,  we allow for
a nonzero radial acceleration.

For a purely spherical expansion with radial velocity $v$ and radial
velocity gradient $dv/dr$, 
let us first define an expansion anisotropy factor,
\begin{equation}
\sigma \equiv \frac{r}{v} \frac{\partial v}{\partial r} - 1 
\, .
\label{eqn:sigma}
\end{equation}
The Sobolev optical depth along a direction $\cos\, \mu$ can then be written 
in the form
\begin{equation}
\tau_\mu = \frac{\tau_0}{1 + \sigma \mu^2}
\, ,
\label{eqn:taumu}
\end{equation}
where $\tau_{0}$ is a characteristic optical depth in the lateral
($\mu=0$) direction [as defined further in eqn. (\ref{eqn:tau0atomic})
below].
The angle-dependent Sobolev escape probability is then given by
\begin{equation}
p ( \mu ) = \frac{1 - e^{-\tau_\mu}}{\tau_\mu}
\, .
\end{equation}
For line photons trapped within a local Sobolev resonance layer,
this represents the probability of escape {\em per scattering}.
When normalized  by the angle-averaged escape probability,
\begin{equation}
\pavg \equiv \frac{1}{2}
\int_{-1}^1 p(\mu ) \, d\mu
\, ,
\end{equation}
then $p(\mu)/\pavg$ gives the relative angle
distribution of net line emission from a localized site of thermal
photon creation.

To include the effects of resonance scattering on X-ray line emission,
we can thus generalize the basic OC01 formalism, developed for the
purely optically thin emission, by simply multiplying the local emissivity
[cf. eqn.~(\ref{eqn:lx}) or OC01 eqn. (9)] by this normalized angle
distribution:
\begin{equation}
L_x = L_0 \int^{u_x}_{0} du \frac{f_X(u)}{w^3(u)} \left[ e^{-\tau_c(\mu,u)}
\frac{p(\mu)}{\pavg} \right]_{\mu = -x/w(u)}
\, .
\end{equation}
For optically thin lines ($\tau_{\mu} \ll  1$), 
the angle emission correction 
recovers the simple isotropic scaling 
$ p(\mu)/\left < p \right >  = 1$,
while for optically thick lines ($\tau_{\mu} \gg 1$), 
it takes the form
\begin{equation}
\frac{p(\mu)}{\pavg} =  
\frac{1+ \sigma \mu^{2} }{1+\sigma/3}
\, .
\end{equation}
The analysis by IG02 examines this case of optically thick line
emission in a constant speed expansion, for which $dv/dr=0$ implies
$\sigma = -1$ and thus
\begin{equation}
\frac{p(\mu)}{\pavg} =  
    \frac{3}{2} \, (1 - \mu^{2})
\, .
\end{equation}

To allow for a non zero radial velocity gradient, let us use here a
similar $\beta$-law velocity form to compute the  anisotropy factor defined 
in eqn. (\ref{eqn:sigma}),
\begin{equation}
\sigma =  \frac{\bsob u}{1 - u} - 1
\, .
\label{eqn:sigbeta}
\end{equation}
Note that this is independent of the wind terminal speed 
and for a given radius depends only on the velocity index $\bsob$.
Taking $\bsob = 0$ recovers the constant expansion case $\sigma = -1$, 
while $\bsob > 0$ gives $\sigma > -1$ at the lower radii ($u > 0$), where
the wind is still accelerating.
Overall, the parameter $\bsob$ thus provides a convenient proxy for
varying the relative importance of flow acceleration (compared to spherical
divergence) in the local Sobolev escape scalings of the X-ray emission.

Our introduction here of a separate symbol, $\bsob$, for the velocity index
relevant for Sobolev escape reflects the notion that in such a
complex flow the local regions of X-ray emission need not always rigorously
follow the global velocity of the bulk wind outflow.
In particular, since the X-ray emitting material is generally too
highly ionized to be directly driven by line-opacity, it might
feasibly be better modeled as having a locally flat velocity gradient,
$dv/dr=0$, which would then be represented by a velocity index
$\bsob=0$ instead of the $\beta \approx 1$ used to model the overall
wind outflow.
On the other hand, it might also be that the X-ray emitting gas is
hydrodynamically so tightly coupled to the mean wind that even on a 
local scale of resonance trapping, it too exhibits a similarly positive 
acceleration that is best represented by taking $\bsob = \beta = 1$.
In the line-fitting analysis below, we thus consider both these
options.

Instead of the IG02 assumption of optically thick lines,
our analysis also allows general parameterization of the  line optical
thickness, as set by the overall factor $\tau_0$ that gives
the Sobolev optical depth in the lateral 
direction $\mu = 0$.
In terms of fundamental atomic parameters, this is given by
\begin{equation}
\tau_0 = \frac{3}{8}\, \frac{\lambda}{r_e}\, \frac{c}{v}\, f_i\, n_i\,
\sigma_T\, r. 
\label{eqn:tau0atomic}
\end{equation}
Here $r_e$ is the classical electron radius, $\sigma_T$ is the Thomson
cross section, $f_i$ is the oscillator strength of the transition, and $n_i$
is the ion density. Using the steady state mass loss rate
$\dot{M} = 4 \pi \rho r^2 v$ within the assumed $\beta$ velocity law,
this can be recast in a form explicitly showing the dependence on
inverse radius $u$,
\begin{equation}
\tau_0 = \tss \frac{u}{w^2(u)}
\, ,
\label{eqn:tau0}
\end{equation}
where we have defined a characteristic Sobolev optical depth,
\begin{equation}
\tss = \frac{\lambda r_e c \dot{M}}{4 R_* v^2_\infty} 
\left( f_i \frac{n_i}{\rho}\right)
\, .
\end{equation}
The factor 
\begin{equation}
\frac{n_i}{\rho} = \frac{n_i}{n_e}\, \frac{n_e}{\rho} = 
\frac{A_i\, q_i\, f_X}{\mu_N\, m_p}
\end{equation}
gives the ratio of the ion number density to the mass density. Here $A_i$ is
the abundance of the element relative to hydrogen, $q_i$ is the ion fraction,
$f_X$ is the filling factor of X-ray emitting plasma, and $\mu_N m_p$ is the
mean mass per particle. We take this ratio to be a constant with radius,
although in principle the ion fraction and filling factor could vary.

In this paper, we take \tss\ as a free parameter. For general values of the
optical depth, the angle-averaged escape probability cannot be evaluated
analytically. Fortunately, \citet[pp.128--129]{Castor04} gives a
computationally efficient approximation (attributed to G. Rybicki) that is
accurate to $\sim 1.5\%$, so we use this approximation to calculate $\pavg$
for finite values of \tss.

We have implemented this X-ray emission formalism as a local model in XSPEC. 
The Sobolev optical depth has angular and radial dependence as given by
eqn.~(\ref{eqn:taumu}) and (\ref{eqn:tau0}). 
The additional parameters added to the OC01 model are a
switch to turn on or off completely optically thick scattering; the
characteristic Sobolev optical depth \tss\ (used when the
completely optically thick switch is off); and the value of the velocity law
exponent used in calculating $\sigma$, \bsob.

\begin{figure}[p]
  \begin{center}
    \plotone{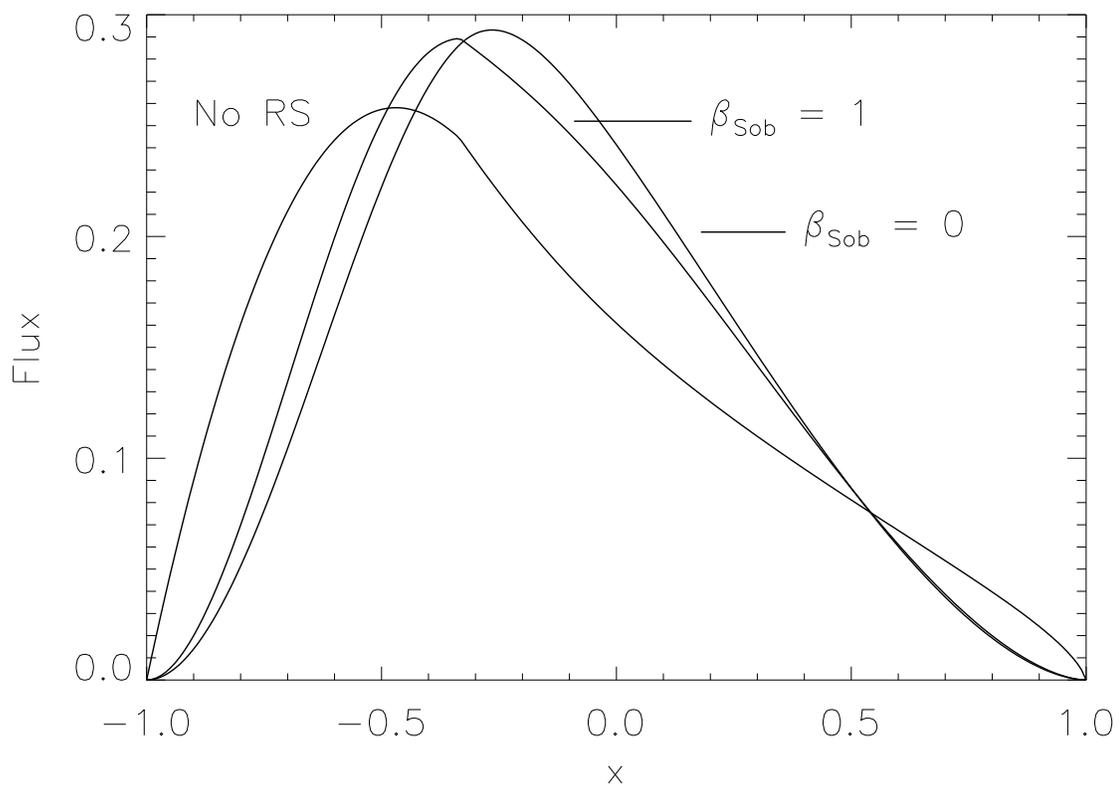}
  \end{center}
\caption{Comparison of the influence of different values of \bsob\ on Doppler
  profile shape. All models have $q = 0$, $u_0 = \frac{2}{3}$, and $\tau_* =
  5$. The most asymmetric model is optically thin. Both of the other models
  use the approximation that \tss\ is infinite; the more asymmetric of the two
  has $\bsob = 1$, while the least asymmetric has $\bsob = 0$.} 
\label{fig:rsnors}
\end{figure}

\begin{figure}[p]
  \begin{center}
    \plotone{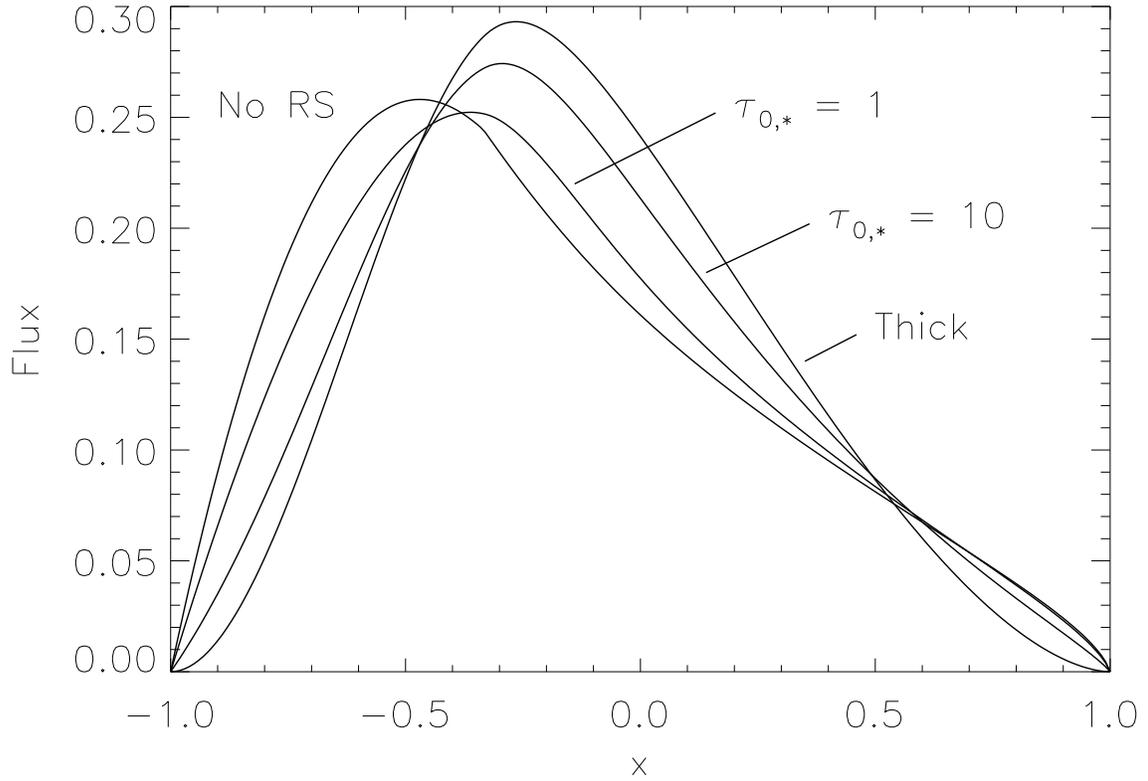}
  \end{center}
\caption{Comparison of the influence of various values of the characteristic
  Sobolev optical depth \tss\ on Doppler profile shape. All models have
  $q = 0$, $u_0 = \frac{2}{3}$, $\tau_* = 5$, and $\bsob = 0$. In order from
  most asymmetric to least, the models have $\tss = 0, 1, 10,$ and $\infty$.}  
\label{fig:rsprof}
\end{figure}

Figures~\ref{fig:rsnors} and \ref{fig:rsprof} compare sample results
for X-ray line profiles assuming various values of \tss\ and \bsob.
The overall trend is for higher values of \tss\ and lower values of \bsob\ to
give more symmetric profiles. The trend of lower values of \bsob\ to give more
symmetric profiles is what one would expect; lowering \bsob\ suppresses the
effect of the flow acceleration in promoting radial photon escape, thus
enhancing lateral escape an symmetrizing the profile.

\subsection{Best-fit model including resonance scattering \label{sec:rsfit}}

\begin{figure}[p]
  \begin{center}
    \plotone{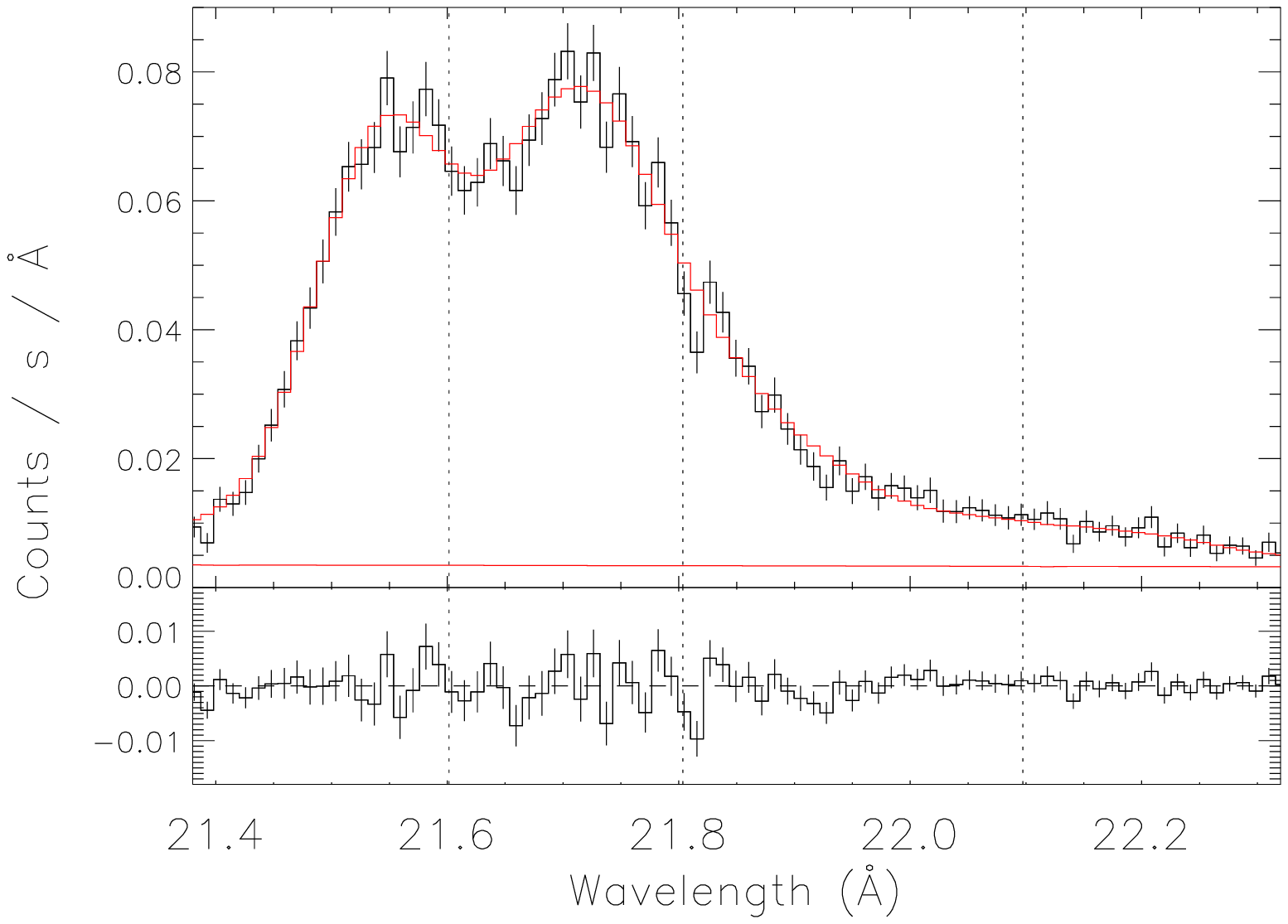}
  \end{center}
\caption{\ion{O}{7} triplet with best-fit model assuming resonance scattering
  with $\bsob = 1$. Scheme is as in Fig.~\ref{fig:o7nors}.}
\label{fig:o7rsb1}
\end{figure}

\begin{figure}[p]
  \begin{center}
    \plotone{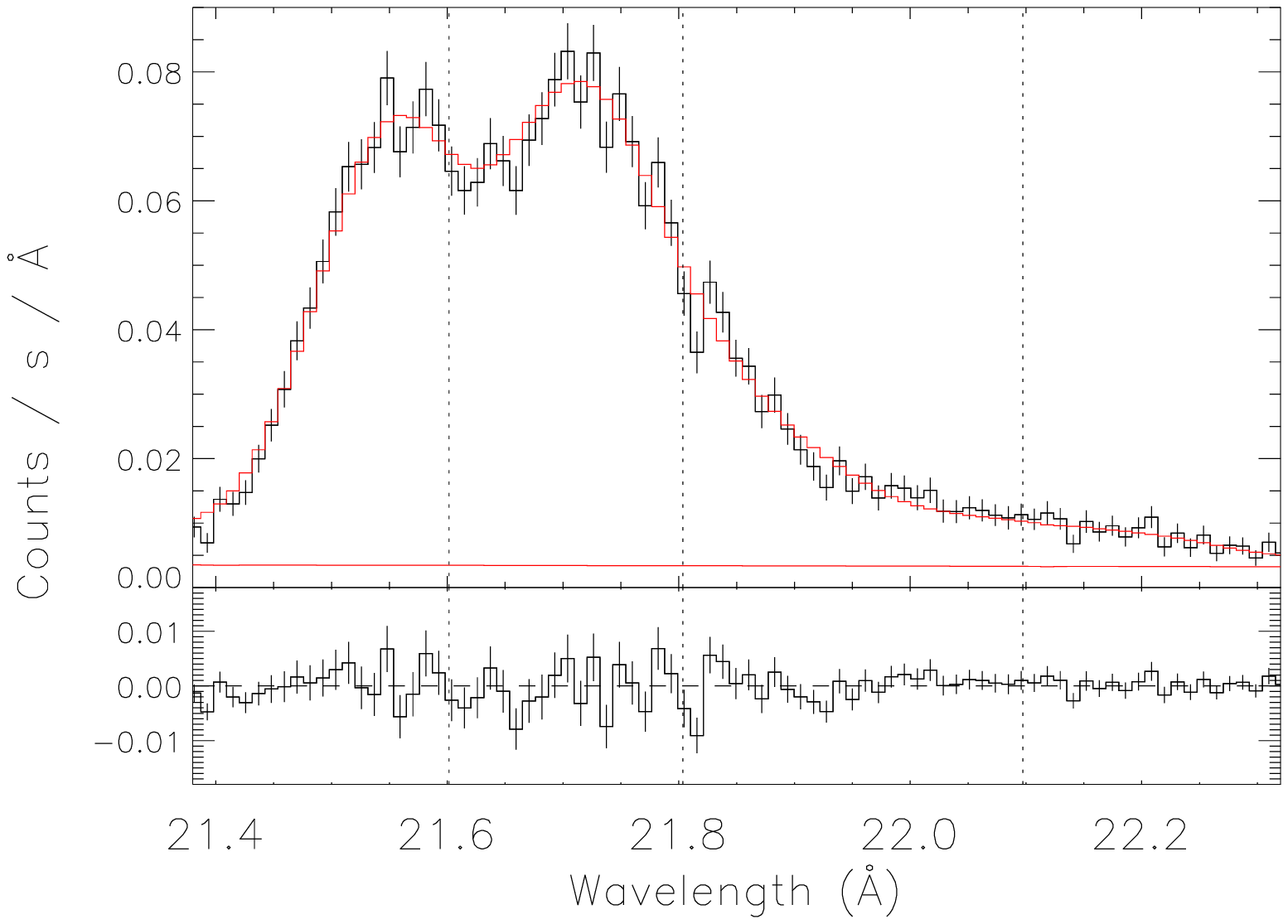}
  \end{center}
\caption{\ion{O}{7} triplet with best-fit model assuming resonance scattering
  with $\bsob = 0$. Scheme is as in Fig.~\ref{fig:o7nors}.}
\label{fig:o7rsb0}
\end{figure}

\begin{figure}[p]
  \begin{center}
    \plotone{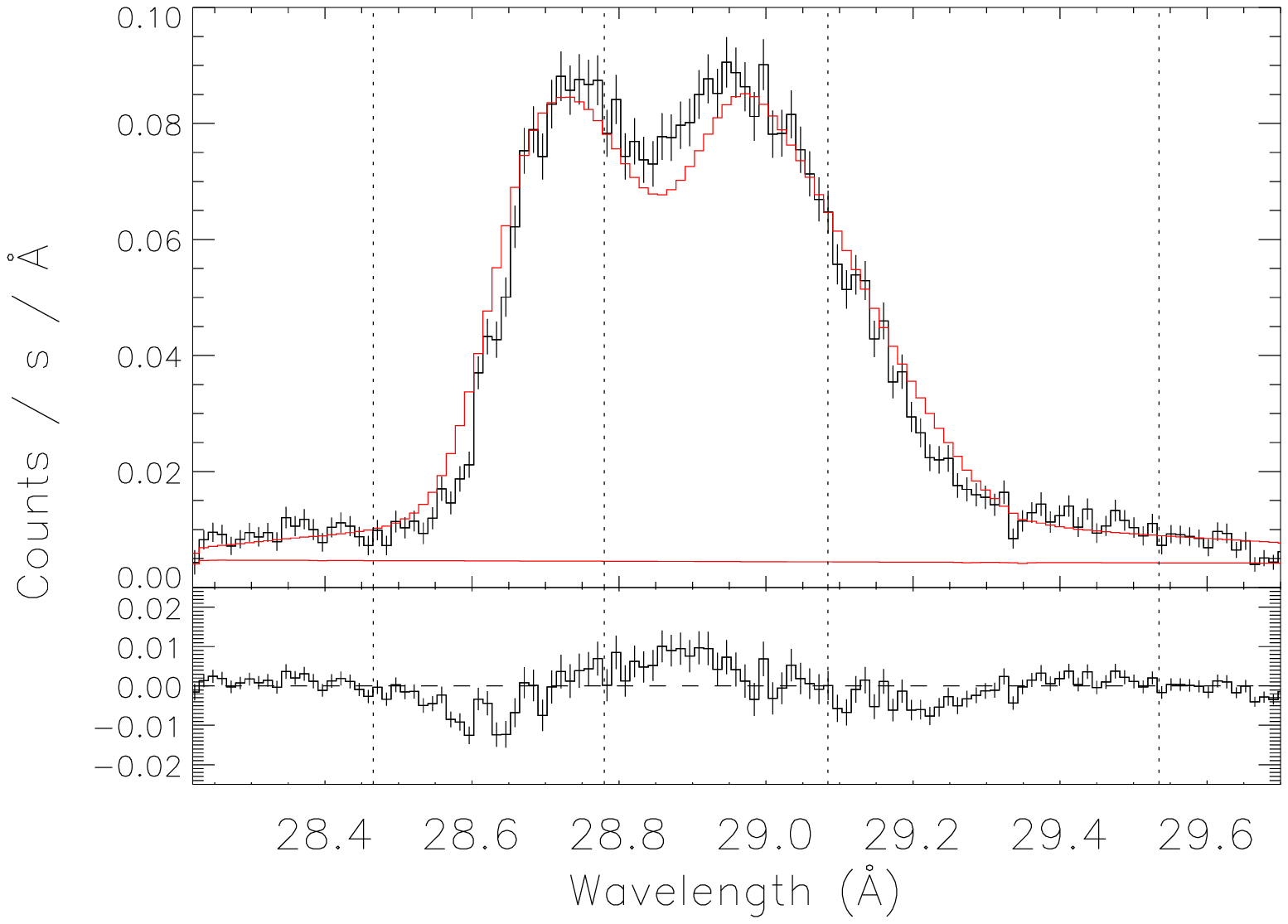}
  \end{center}
\caption{\ion{N}{6} triplet with best-fit model assuming resonance scattering
  with $\bsob = 1$. Scheme is as in Fig.~\ref{fig:o7nors}.}
\label{fig:n6rsb1}
\end{figure}

\begin{figure}[p]
  \begin{center}
    \plotone{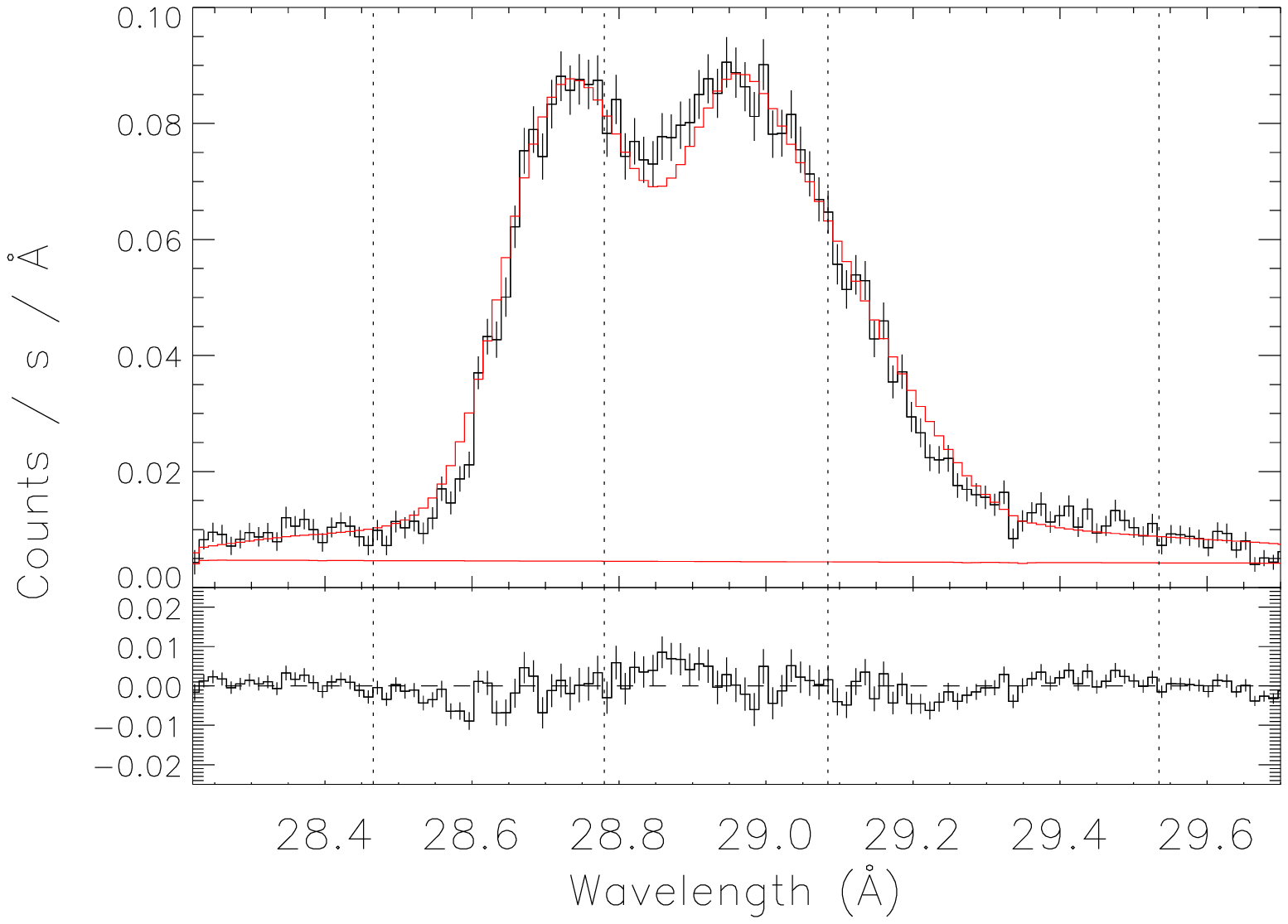}
  \end{center}
\caption{\ion{N}{6} triplet with best-fit model assuming resonance scattering
  with $\bsob = 0$. Scheme is as in Fig.~\ref{fig:o7nors}.}
\label{fig:n6rsb0}
\end{figure}

In this section we fit He-like profile models including resonance scattering
to the \ion{O}{7} and \ion{N}{6} complexes. We fit each complex twice: once
assuming $\bsob = 1$ and once assuming $\bsob = 0$. The best-fit models are
shown in Figures \ref{fig:o7rsb1}, \ref{fig:o7rsb0}, \ref{fig:n6rsb1}, and
\ref{fig:n6rsb0}. The best-fit parameters are given in Tables~\ref{tab:O7} and
\ref{tab:N6}. 

The \ion{O}{7} profile is well fit by either value of \bsob. We tested
goodness of fit by comparing the fit statistic of 1000 Monte Carlo
realizations of the model to the fit statistic of the data; both models are
formally acceptable. The fit with $\bsob = 1$ is better than that with $\bsob
= 0$ , but only by $\Delta C = 3.8$, which is about $2\sigma$ for one
interesting parameter. The fit with $\bsob = 0$ has a
significantly smaller value of \tss\ than the fit with
$\bsob = 1$, as would be expected. The fit with
$\bsob = 1$ is statistically consistent with the
approximation that the Sobolev optical depth becomes infinite.

The \ion{N}{6} profile is much better fit by either model including resonance
scattering than it is by the original model. Furthermore, the model with
$\bsob = 0$ gives a significantly better fit than the model
with $\bsob = 1$. However, neither model is formally acceptable,
and even the $\bsob = 0$ model shows residuals of the same
qualitative form as the original model, albeit of a much lower strength. For
both models including resonance scattering, the optically thick approximation
gives a better fit than a profile with finite Sobolev optical depth.

To test the significance of profile broadening introduced by co-adding data
with random systematic errors in the wavelength scale, we have also fit each
best-fit model with an additional 7 m\AA\ Gaussian broadening. In all cases,
the best-fit parameters did not change significantly and the fit statistics
were not significantly worse. Thus we conclude that our analysis is not
strongly affected by this broadening.

\section{Discussion \label{sec:discussion}}

\subsection{Comparison of results}

The profile fits presented in \S~\ref{sec:nrfit} clearly show that the
\ion{O}{7} and \ion{N}{6} He-like triplet complexes in \zp\ cannot be fit by
models that assume the same profile shapes for the resonance and
intercombination lines. The profile fits presented in \S~\ref{sec:rsfit} show
that these complexes can be much better fit by a model including the effects
of resonance scattering.

However, although the \ion{O}{7} complex is well fit by a model including the
effects of resonance scattering, the \ion{N}{6} complex shows differences in
profile shape between the resonance and intercombination line that are greater
than our model can reproduce, even under the most generous conditions
($\tss \rightarrow \infty, \bsob = 0$).
Furthermore, one would expect the two complexes to show relatively similar
parameters; for example, since the elemental abundance of nitrogen appears to
be roughly twice that of oxygen, one would expect the parameter
\tss\ to be about twice as large for the fit to \ion{N}{6} as it is
for \ion{O}{7}. But a fit to the \ion{N}{6} profile with
$\bsob = 0$ and $\tss \approx 10$ (roughly twice the
value measured for \ion{O}{7}) would give a substantially  worse fit than a
model with infinite Sobolev optical depth, which itself has significant
residuals. 

The fact that the apparent discrepancy between the shapes of the resonance
and intercombination line profiles is much greater for \ion{N}{6}
than for \ion{O}{7} implies that whatever the symmetrizing mechanism for the
resonance line is, it is significantly stronger for \ion{N}{6}. There is no
obvious explanation for this in the resonance scattering paradigm.

\subsection{Plausibility of the importance of resonance scattering}

It is worth revisiting the plausibility arguments of IG02 to confirm that one
would expect resonance scattering to be important for these ions in the wind
of \zp. The relevant quantities to estimate are the Sobolev optical depth and
the ratio of the Sobolev length to the cooling length.

The Sobolev length is given by \citep[e.g.][]{G95}
\begin{equation}
L_\mu = \frac{1 + \sigma}{1 + \sigma \mu^2} \frac{v_{\mathrm{th}}}{dv/dr} =
\frac{v_{\mathrm{th}}}{v/r} \frac{1}{1 + \sigma \mu^2}\, .
\end{equation}
The cooling length is given by
\begin{equation}
\frac{5}{2} \frac{k \Delta T}{n_e \lambda} v,
\end{equation}
as derived in IG02.

Taking the ratio,
\begin{eqnarray}
\frac{L_c}{L_\mu} &=& \frac{5}{2} \frac{k \Delta T}{n_e \Lambda}
\frac{v}{v_{th}} \frac{v}{r} (1 + \sigma \mu^2) \\
&=& \frac{5}{2} \frac{k \Delta T}{\Lambda} \frac{4 \pi \mu_N m_p}{\dot{M}}
\frac{v_\infty}{v_{th}} v_\infty^2 R_* \frac{w^3(u) f_X}{u} (1 + \sigma \mu^2)
\end{eqnarray}
where we have used $\dot{M} = 4 \pi \mu_N m_p n_e r^2 v$ for a smooth wind,
and added a filling factor $f_X$ to correct for the ratio of the density of
the X-ray emitting plasma to the mean density expected for a smooth wind.

Putting in some representative numbers appropriate to \zp, we have
\begin{equation}
\frac{L_c}{L_\mu} = 10\, (1 + \sigma \mu^2)\, \frac{w^3(u) f_X}{u}\, 
\frac{1}{\dot{M}_6}
\end{equation}
where $\dot{M}_6$ is the mass-loss rate in units of $10^{-6} \,
\mathrm{M_\odot\, yr^{-1}}$. We have used $\Lambda = 6 \times 10^{23}\,
\mathrm{ergs\, s^{-1}\, cm^3}$, $\Delta T = 2 \mathrm{MK}$, $\mu_N = 0.6$,
$v_{th} = 50\, \mathrm{km\, s^{-1}}$, $v_\infty = 2500\, \mathrm{km\,
  s^{-1}}$, and $R_* = 1.4 \times 10^{12}\, \mathrm{cm}$.

This expression is greater than unity for lateral escape except at
small radii ($r < 2 R_*$) if the filling factor is of order unity. However,
if the filling factor is significantly less than unity, the Sobolev
approximation may not be valid.


We now consider the expected values of the characteristic Sobolev optical
depth,
\begin{equation}
\tss = \frac{\lambda\, r_e\, c\, \dot{M}}{4 R_*\, v^2_\infty} \left( f_i
\frac{n_i}{\rho}\right)
= \frac{\lambda\, r_e\, c\, \dot{M}}{4 \mu_N\, m_p\, R_*\, v^2_\infty} 
f_i\, A_i\, q_i\, f_X.
\label{eqn:tau0star}
\end{equation}

Putting in appropriate values, we get
\begin{equation}
\tss = 120\, 
\left(f_i\, \frac{A_i}{10^{-3}} \frac{\lambda}{20\, \mbox{\AA}}\right)\, 
q_i\, f_X\, \dot{M}_6
\label{eqn:calctau0star}
\end{equation}

We give calculations of $\tss / q_i f_X$ for important lines in O star
spectra in Table~\ref{tab:tau0star}. We have assumed solar abundances for all
elements except C, N, and O \citep{AG89}. We assumed that the sum of CNO is
equal to the solar value, with carbon being negligible and with nitrogen
having twice the abundance of oxygen; this is an estimate based on the
observed X-ray emission line strengths. Note that the Sobolev optical depth
scales with the wavelength of the transition; this means that the Sobolev
optical depths are significantly smaller for an X-ray transition than they are
for a comparable UV transition. It also means that the longer wavelength
K-shell transitions of N and O will be more strongly affected by resonance
scattering than the shorter wavelength K-shell transitions of Ne, Mg, and Si,
and L-shell transitions of Fe, a trend that is reinforced by the high
elemental abundances of N and O.

Again, if the X-ray filling factors are of order unity, the characteristic
Sobolev optical depths for the resonance lines of \ion{N}{6} and \ion{O}{7}
are large, but X-ray filling factors of order $10^{-3}$ or less are
sufficient to cause the lines to become optically thin. However, the
requirement that the Sobolev length in the lateral direction be smaller than
the cooling length is about as stringent, so if resonance scattering is
important for strong lines, the Sobolev approximation should also be valid. 

The high filling factors required are at odds with the simple two-component
fluid picture of the OC01 model, since the X-ray filling factors are known to be
very low. However, if we take the wind to be resolved into the two components
on scales of the order of the Sobolev length, the filling factor would just be
ratio of the local density to the mean density at that radius. This filling
factor would still likely be less than unity for the X-ray emitting plasma,
but not as low as the X-ray filling factor for the whole wind. This conjecture
is a significantly stronger assumption than is made in OC01. 

\subsection{Impact of resonance scattering on Doppler profile model parameters}

If resonance scattering is important in Doppler profile formation in the X-ray
spectra of O stars, it may lead to a partial reconciliation with the
literature mass-loss rates. The best-fit models for \ion{O}{7} have $\tau_* =
4.1$, and the best-fit model for \ion{N}{6} has $\tau_* = 3.0$. If 
we speculate that somehow the resonance line of \ion{N}{6} is even further
symmetrized than predicted by our model, as the residuals in our best-fit
model imply, the value of $\tau_*$ demanded by the intercombination line
profile residuals should be somewhat higher; a reasonable guess would be
$\tau_* \sim 4 - 5$.

These characteristic optical depths are higher than those measured by
\citet{KCO03} for \zp\ by applying the model of OC01 to Doppler profiles
observed with the {\it Chandra} HETGS; the lines studied in that paper were
mostly resonance lines as well. They are still somewhat lower than one
would expect given the literature mass-loss rates; however, a detailed
comparison with opacity calculations and mass-loss rates remains to be
done. Furthermore, new, sophisticated analyses of UV absorption line profiles
indicate that the published mass-loss rates of O star winds are too high
\citep{MFSH03, Hel03, BLH05}; the most recent systematic analysis of Galactic
O stars finds that for at least some spectral types, the published mass-loss
rates must be at least an order of magnitude too great \citep{FMP06}, but even
the more conservative revisions lower the mass-loss rates by a factor of a
few.

The fact that our measurements have $q \sim 0$ for both \ion{O}{7} and
\ion{N}{6} is an important additional constraint. In cases in which one would
like to fit a single emission line, it is desirable to have as few free
parameters as possible. If we can assume $q = 0$, we reduce the fitting of the
profile shape to two free parameters for a line with no resonance scattering
($R_0$ and $\tau_*$) and three free parameters for a line with resonance
scattering (the previous parameters in addition to \tss). Furthermore, for
lines where $\tau_*$ is large enough to obscure the inner part of the wind,
$R_0$ is effectively removed as a free parameter, further reducing the number
of free parameters to one and two for non-resonance and resonance lines,
respectively. Thus, high signal-to-noise ratio Doppler profiles with
significant continuum absorption and no resonance scattering may provide robust
measurements of the mass-loss rates of O stars. A good candidate for this is
the 16.78 \AA\ line of \ion{Fe}{17}, which is likely not to be optically
thick, and which is not blended with other lines.

\subsection{Future work}

Here we give a list of issues raised by this analysis that should be
addressed in future work.

1. The discrepancies in the fits in this paper must be resolved. The fact that
we cannot fit the \ion{N}{6} profile well is unsatisfactory. The difference
between the appearance of the \ion{N}{6} complex and the \ion{O}{7} complex
requires explanation.

2. The effect of resonance scattering on other resonance lines in the X-ray
spectrum should be considered. Furthermore, unless we can make concrete
predictions for the importance of resonance scattering for these lines, there
may be significant fitting degeneracies between resonance scattering and low
characteristic continuum optical depths.

3. The effect of multiple lines on resonance scattering should be explored. Of
special importance is the calculation of the profile of a close doublet, such
as \lya. In that case, the splitting between the two lines is of the order of
the thermal velocity of the ions. 

\section{Conclusions \label{sec:conclusions}}

We have fit Doppler profile models based on the parametrized model of
OC01 to the He-like triplet complexes of \ion{O}{7} and \ion{N}{6} in
the high signal-to-noise ratio \xmm\, RGS X-ray spectrum of \zp. We find
that the complexes cannot be well fit by models assuming the same shape for
the resonance and intercombination lines; the predicted resonance lines are
too blue and the predicted intercombination lines are too red. This effect is
what is predicted qualitatively if resonance scattering is important.

We find that models including the effects of resonance scattering give
significantly better fits. However, there is significant disagreement between
the \ion{O}{7} and \ion{N}{6} profiles in the degree of resonance line
symmetrization that is difficult to understand in the framework of the
resonance scattering model. Nevertheless, the general trend of the resonance
scattering model to give more symmetrized profiles provides an interesting
alternative (or supplement) to models that assume reduced wind attenuation due
to reduced mass-loss rates and/or porosity.

\acknowledgements
We acknowledge comments from an anonymous referee that helped clarify the
presentation of this paper. We acknowledge useful conversations with David
Cohen. We thank David Cohen and Dave Spiegel for their careful reading of the
manuscript. M.A.L. acknowledges NASA grant NNG04GL76G.


\bibliography{/Users/maurice/papers/xray-ostar}

\clearpage

\begin{deluxetable}{ccc}
  \tablewidth{0pt}
  \tablecaption{List of observations with net exposure times \label{tab:obslog}}
  \tablehead{
    \colhead{} &
    \colhead{$t_{\mathrm{exp, RGS1}}$ \tablenotemark{b}} &
    \colhead{$t_{\mathrm{exp, RGS2}}$ \tablenotemark{b}} \\
    \colhead{ObsID \tablenotemark{a}} & 
    \colhead{(ks)} &
    \colhead{(ks)}
  }
  \startdata
  0095810301 & 30.6 & 29.8 \\
  0095810401 & 39.7 & 38.3 \\
  0157160401 & 41.5 & 40.2 \\
  0157160501 & 32.8 & 32.8 \\
  0157160901 & 43.4 & 43.4 \\
  0157161101 & 27.0 & 27.0 \\
  0159360101 & 59.2 & 59.2 \\
  0159360301 & 22.0 & 22.0 \\
  0159360501 & 31.5 & 31.5 \\
  0159360901 & 46.6 & 46.6 \\
  0159361101 & 41.1 & 41.0 \\
  \enddata
  \tablenotetext{a}{\xmm\, Observation ID.}
  \tablenotetext{b}{Net exposure time.}
\end{deluxetable}

\begin{deluxetable}{cccccccccc}
  \tablewidth{0pt}
  \tablecaption{Model fit parameters for \ion{O}{7} \label{tab:O7}}
  \tablehead{
    \colhead{Resonance} &
    \colhead{} &
    \colhead{} &
    \colhead{} &
    \colhead{} &
    \colhead{} &
    \colhead{} &
    \colhead{} &
    \colhead{} &
    \colhead{} \\
    \colhead{Scattering} &
    \colhead{\bsob} &
    \colhead{$q$} & 
    \colhead{$\tau_*$} &  
    \colhead{$u_0$} &
    \colhead{\tss} &
    \colhead{$G$ \tablenotemark{a}} &
    \colhead{$n$ \tablenotemark{b}} &
    \colhead{$C$ \tablenotemark{c}} &
    \colhead{MC \tablenotemark{d}}
  }
  \startdata
  No & \nodata & -0.21             & 1.6               & 0.62     & 
  \nodata         & 0.91                 & 6.91            & 152.2 & \nodata \\
  Yes & 1   & $0.15^{+0.06}_{-0.07}$ & $4.1^{+0.3}_{-0.4}$ & $> 0.68$ & 
  $> 50$          & $1.11^{+0.03}_{-0.04}$   & $6.88 \pm 0.07$ & 85.3 & 0.578 \\
   & 0      & $0.15 \pm 0.07$ & $4.1 \pm 0.4$     & $> 0.63$ & 
  $5.9^{+3.2}_{-1.8}$ & $1.02^{+0.04}_{-0.03}$ & $6.88 \pm 0.07$ & 89.1 & 0.730 
  \enddata
  \tablenotetext{a}{$G = (f + i) / r$ is assumed not to vary with radius.}
  \tablenotetext{b}{Normalization of entire complex ($r+i+f$) in units of
    $10^{-4}$ photons $\mathrm{cm^{-2}\, s^{-1}}$.}
  \tablenotetext{c}{For 83 bins.}
  \tablenotetext{d}{Fraction of 1000 Monte Carlo realizations of model having
    $C$ less than the data.}
  \tablecomments{The first row gives the best fit for a model not including
    resonance scattering (i.e. the model of OC01 and Leutenegger et al.). The
    second row gives the best fit for a model including resonance scattering
    with $\bsob = 1$, and the last row has $\bsob= 0$. We used a value of $P =
    1.67 \times 10^4$ for all \ion{O}{7} profile models \citep{LPKC06}.} 
\end{deluxetable}

\begin{deluxetable}{cccccccccc}
  \tablewidth{400pt}
  \tablecaption{Model fit parameters for \ion{N}{6} \label{tab:N6}}
  \tablehead{
    \colhead{Resonance} &
    \colhead{} &
    \colhead{} &
    \colhead{} &
    \colhead{} &
    \colhead{} &
    \colhead{} &
    \colhead{} &
    \colhead{} &
    \colhead{} \\
    \colhead{Scattering} &
    \colhead{\bsob} &
    \colhead{$q$} & 
    \colhead{$\tau_*$} &  
    \colhead{$u_0$} &
    \colhead{\tss} &
    \colhead{$G$ \tablenotemark{a}} &
    \colhead{$n$ \tablenotemark{b}} &
    \colhead{$n_{\beta}$ \tablenotemark{c}} &
    \colhead{$C$ \tablenotemark{d}}
  }
  \startdata
  No  & \nodata & -0.34 & 0.5 & 0.58 & \nodata & 0.87 & 1.562 & 0 & 510.5 \\
  Yes & 1       & -0.09 & 2.1 & 0.50 & thick   & 1.10 & 1.559 & 0.87 & 292.2 \\
      & 0       & 0.06 & 3.0 & 0.48 & thick   & 1.15 & 1.552 & 1.25 & 188.4
  \enddata
  \tablenotetext{a}{$G = (f + i) / r$ is assumed not to vary with radius.}
  \tablenotetext{b}{Normalization of entire \ion{N}{6} complex in units of
    $10^{-3}$ photons $\mathrm{cm^{-2}\, s^{-1}}$.}
  \tablenotetext{c}{Normalization of \ion{C}{6} \lyb\ in units of
    $10^{-5}$ photons $\mathrm{cm^{-2}\, s^{-1}}$.}
  \tablenotetext{d}{For 117 bins.}
  \tablecomments{The first row gives the best fit for a model not including
    resonance scattering (i.e. the model of OC01 and Leutenegger et al.). The
    second row gives the best fit for a model including resonance scattering
    with $\bsob = 1$, and the last row has $\bsob = 0$. The \ion{C}{6} \lyb\
    line is assumed to have the same values of $q$, $\tau_*$, and $u_0$ as the
    \ion{N}{6} triplet, and is assumed not to be affected by resonance
    scattering. We used a value of $P = 1.01 \times 10^5$ for all \ion{N}{6}
    profile models \citep{LPKC06}.} 
\end{deluxetable}

\begin{deluxetable}{cccccc}
  \tablewidth{0pt}
  \tablecaption{Expected characteristic Sobolev optical depth \label{tab:tau0star}}
  \tablehead{
    \colhead{} &
    \colhead{} &
    \colhead{$\lambda$ \tablenotemark{a}} &  
    \colhead{$f_i$ \tablenotemark{b}} &
    \colhead{$A_i$ \tablenotemark{c}} &
    \colhead{$\tss / q_i\, f_X$ \tablenotemark{d}}\\
    \colhead{Ion} &
    \colhead{Transition} &
    \colhead{(\AA)} &
    \colhead{} &
    \colhead{($10^{-3}$)} &
    \colhead{}
  }
  \startdata
  \ion{N}{6} & r       & 28.78 & 0.6599 & 0.9   & 103 \\
  & $\beta$            & 24.90 & 0.1478 &       & 20 \\
  \ion{N}{7} & \lya   & 24.78 & 0.1387, 0.2775 \tablenotemark{e} & & 19, 37 \\
  \ion{O}{7} & r       & 21.60 & 0.6798 & 0.45  & 40 \\
  & $\beta$            & 18.63 & 0.1461 &       & 7 \\
  \ion{O}{8} & \lya   & 18.97 & 0.1387, 0.2775 \tablenotemark{e} & & 7, 14 \\
  \ion{Fe}{17} &       & 15.01 & 2.517  & 0.047 & 11 \\
  &                    & 15.26 & 0.5970 &       & 2.5 \\
  &                    & 16.78 & 0.1064 &       & 0.5 \\
  &                    & 17.05 & 0.1229 &       & 0.6 \\
  \ion{Ne}{9} & r      & 13.45 & 0.7210 & 0.12  & 7.0 \\
  & $\beta$            & 11.55 & 0.1490 &       & 1.2 \\
  \ion{Ne}{10} & \lya & 12.13 & 0.1382, 0.2761 \tablenotemark{e} & & 1.2, 2.4 \\
  \ion{Mg}{11} & r     & 9.17  & 0.7450 & 0.038 & 1.6 \\
  \ion{Mg}{12} & \lya & 8.42  & 0.1386, 0.2776 \tablenotemark{e} & & 0.27, 0.53\\
  \ion{Si}{13} & r     & 6.65  & 0.7422 & 0.036 & 1.1 \\
  \ion{Si}{14} & \lya & 6.18  & 0.1386, 0.2776 \tablenotemark{e} & & 0.19, 0.37 \\

  \enddata
  \tablenotetext{a}{Wavelength.}
  \tablenotetext{b}{Oscillator strengths are from CHIANTI \citep{Del97,
      Landi2006}.}
  \tablenotetext{c}{Assumed abundance relative to hydrogen.} 
  \tablenotetext{d}{This number is calculated using
    eqn.~(\ref{eqn:calctau0star}) assuming a mass-loss rate of $10^{-6}\,
    \mathrm{M_{\odot}\, yr^{-1}}$.}
  \tablenotetext{e}{\,\lya\ is a doublet.}
\end{deluxetable}

\clearpage

\clearpage

\end{document}